\documentclass[twocolumn]{aastex631}

\newcommand{\kzz}{$K_\textrm{zz}$}
\usepackage{float}
\usepackage{comment}
\usepackage{amsmath, amssymb, amsthm}
\usepackage{booktabs}
\usepackage{graphicx}

\begin{document}

\title{Probing the Heights and Depths of Y Dwarf Atmospheres:  A Retrieval Analysis of the JWST Spectral Energy Distribution of WISE J035934.06$-$540154.6}

\author{Harshil Kothari}
\affiliation{Ritter Astrophysical Research Center, Department of Physics \& Astronomy, University of Toledo, 2801 W. Bancroft St., Toledo, OH 43606, USA}

\author{Michael C. Cushing}
\affiliation{Ritter Astrophysical Research Center, Department of Physics \& Astronomy, University of Toledo, 2801 W. Bancroft St., Toledo, OH 43606, USA}

\author{Ben Burningham}
\affiliation{Centre for Astrophysics Research, School of Physics, Astronomy and Mathematics, University of Hertfordshire, Hatfield AL10 9AB}

\author{Samuel A. Beiler}
\affiliation{Ritter Astrophysical Research Center, Department of Physics \& Astronomy, University of Toledo, 2801 W. Bancroft St., Toledo, OH 43606, USA}

\author{J. Davy Kirkpatrick}
\affiliation{IPAC, Mail Code 100-22, Caltech, 1200 E. California Boulevard, Pasadena, CA 91125, USA}

\author{Adam C. Schneider}
\affiliation{United States Naval Observatory, Flagstaff Station, 10391 West Naval Observatory Road, Flagstaff, AZ 86005, USA}

\author{Sagnick Mukherjee}
\affiliation{Department of Astronomy and Astrophysics, University of California, Santa Cruz, 1156 High Street, Santa Cruz, CA 95064, USA}

\author{Mark S. Marley}
\affiliation{Lunar and Planetary Laboratory, University of Arizona, 1629 E. University Boulevard, Tucson, AZ 85721, USA}

\begin{abstract}

We present an atmospheric retrieval analysis of the Y0 brown dwarf WISE J035934.06$-$540154.6 using the low-resolution 0.96--12 $\mu$m JWST spectrum presented in \citet{Beiler_2023}.  We obtain volume number mixing ratios of the major gas-phase absorbers (H$_2$O, CH$_4$, CO, CO$_2$, PH$_3$, and H$_2$S) that are 3--5$\times$ more precise than previous work that used HST spectra.  We also find an order-of-magnitude improvement in the precision of the retrieved thermal profile, a direct result of the broad wavelength coverage of the JWST data.  We used the retrieved thermal profile and surface gravity to generate a grid of chemical forward models with varying metallicity, (C/O)$_\textrm{atm}$, and strengths of vertical mixing as encapsulated by the eddy diffusion coefficient $K_\textrm{zz}$. Comparison of the retrieved abundances with this grid of models suggests that the deep atmosphere of WISE 0359$-$54 shows signs of vigorous vertical mixing with $K_\textrm{zz}=10^9$ [cm$^{2}$ s$^{-1}$].  To test the sensitivity of these results to our 5-knot spline thermal profile model, we performed a second retrieval using the \citet{Madhusudhan_2009} thermal profile model.  While the results of the two retrievals generally agree well, we do find differences between the retrieved values of mass and volume number mixing ratio of H$_2$S with fractional differences of the median values of $-$0.64 and $-$0.10, respectively.  In addition, the 5-knot thermal profile is consistently warmer at pressure between 1 and 70 bar.  Nevertheless, our results underscore the power that the broad-wavelength infrared spectra obtainable with the James Webb Space Telescope have to characterize the atmospheres of cool brown dwarfs.

\end{abstract}


\section{Introduction} \label{sec:introduction}

In the last decade, atmospheric retrieval, a method by which the properties of an atmosphere are inferred directly from an observed spectrum, has become a powerful technique for studying the atmospheres of both brown dwarfs and exoplanets \citep[e.g.,][]{Madhusudhan_2009_retrieval, Line_2014}.  With roots in the study of the planets in our solar system \citep[e.g.,][]{Chahine_1968}, a retrieval determines the thermal profile (i.e. the run of temperature and pressure) and atomic/molecular abundances of an atmosphere by iteratively comparing tens of thousands of model spectra to observations in order to optimize the model parameters.

Previous retrievals of brown dwarfs have mostly focused on the warmer objects that populate the L and T spectral classes \citep{Line_2014, Burningham_2017, Zalesky_2019, Lueber_2022, Adams_2023, Rowland_2023, Vos_2023, Hood_2024}.  These retrievals use relatively broad-wavelength spectra covering a minimum of the 0.8--2.4 $\mu$m wavelength and often extending to 4--5 $\mu$m or even to $\sim 15$ $\mu$m.

The cooler brown dwarfs that populate the Y spectral class are rare (roughly 50 are known), faint ($M_\textrm{H}\gtrsim 21$ mag), and emit-most of their radiation at mid-infrared wavelengths.  As a result, the majority of retrievals that have been performed on them used spectra with limited wavelength coverage and/or signal-to-noise ratio.  \citet{Zalesky_2019} performed  retrievals of 8 Y dwarfs using low-resolution 1--1.7 $\mu$m Hubble Space Telescope spectra \citep{Schneider_2015} and measured the abundances of H$_2$O, CH$_4$, NH$_3$, and upper limits for the abundances of CO and CO$_2$.  The H$_2$O and CH$_4$ abundances were consistent with the predictions of thermochemical equilibrium models, but \citeauthor{Zalesky_2019} suggested that the abundance of NH$_3$ may be affected by vertical mixing within the atmosphere.  Unfortunately the narrow wavelength coverage of the HST spectra limit the precision with which the abundances and the thermal profiles can be measured (uncertainties of $\sim$ 0.14 dex and $\sim$ 200 K, respectively) because they only probe a relatively narrow range of pressures in the atmosphere.  

The launch of the James Webb Space Telescope \citep[hereafter JWST;][]{Gardner_2006} has opened a new frontier in the study of Y dwarfs because low- and moderate-resolution spectra are now available over the 1 to 28 $\mu$m wavelength range.  \citet{Barrado_2023} used several retrieval codes to detect both $^{14}$NH$_3$ and $^{15}$NH$_3$ in the moderate-resolution 4.9--18 $\mu$m spectrum of WISEP J182831.08$+$265037.8 (hereafter WISE 1828$+$25; $T_\textrm{eff}$$\approx$350 K) and found a $^{14}\textrm{N}/^{15}\textrm{N}$ value of $673^{+393}_{-212}$, consistent with formation by gravitational collapse of a molecular cloud.  \citet{Lew_2024} used a moderate-resolution 2.88--5.12 $\mu$m spectrum of WISE 1828$+$26 to obtain abundances of H$_2$O, CH$_4$, CO$_2$, NH$_3$ H$_2$S and measured a C/O value of 0.45$\pm$0.01.

In this paper, we add to the short list of brown dwarf JWST-based retrievals by presenting a retrieval analysis of of WISE J035934.06$-$540154.6 (hereafter WISE 0359$-$54) using the low-resolution 0.96--12 $\mu$m JWST spectrum presented in \citet{Beiler_2023}. WISE 0359--54 has a spectral type of Y0, lies at a distance of 13.57 $\pm$ 0.37 pc \citep[$\varpi_{abs}=73.6 \pm$ 2.0 mas,][]{Kirkpatrick_2021}, and has an effective temperature ($T_\mathrm{eff}$) of 467$^{+16}_{-18}$\, K \citep{Beiler_2023}.   In \S 2, we will briefly discuss the spectrum being used for this analysis. In \S  3, we will discuss the retrieval framework that is used to perform the retrieval analysis for WISE 0359--54. In \S  4, we will present and discuss the retrieved results. Finally, in \S 5,  we will summarize and point out key findings of this retrieval analysis.

\section{The Spectrum} \label{sec:data}

\begin{figure*}[htb!]
    \centering
    \includegraphics[width= \textwidth]{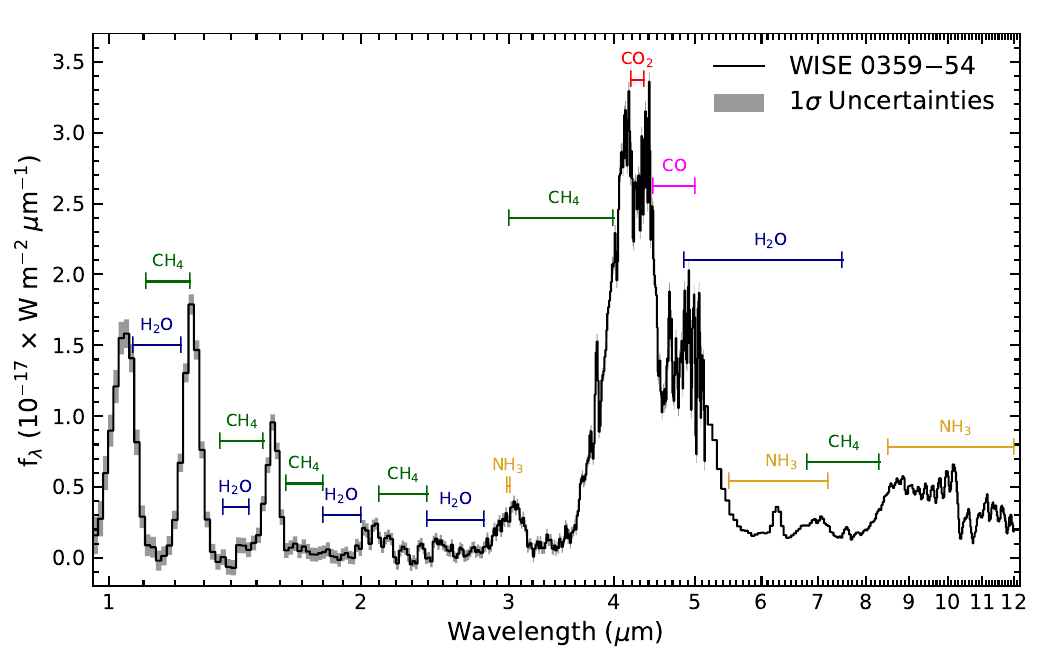}
    \caption{The 0.96--12$\mu$m JWST spectrum with 1$\sigma$ uncertainties (NIRSpec $+$ MIRI LRS) of WISE 0359$-$54 spectrum \citep{Beiler_2023} in units of $f_\lambda$. The typical signal-to-noise of the NIRSpec and MIRI LRS spectra are $\sim$20 and $\sim$100, respectively.  Also plotted are the locations of prominent absorption bands of H$_2$O, CH$_4$, CO, CO$_2$, and NH$_3$.}
   \label{fig:spectrum}
\end{figure*}

We analyzed the 0.96--12 $\mu$m JWST spectrum of the Y0 dwarf WISE 0359--54 presented in \citet{Beiler_2023}.  The spectrum was obtained using the Near Infrared Spectrograph \citep[hereafter NIRSpec,][]{Jakobsen_2022}, which covers 0.6--5.3 $\mu$m, and the Mid-Infrared Instrument \citep[hereafter MIRI,][]{Rieke_2015}, which covers 5--12 $\mu$m.  The resolving power of the spectra are strong functions of wavelength but on average are $R\equiv \lambda / \Delta \lambda \approx 200$.  \citeauthor{Beiler_2023} used Spitzer/IRAC Channel 2 ([4.5]) photometry from \citet{Kirkpatrick_2012} and MIRI F1000W ($\lambda_\textrm{pivot}$= 9.954 $\mu$m) photometry to absolutely flux calibrate the NIRSpec and MIRI spectra to an overall precision of $\sim5\%$.  \citeauthor{Beiler_2023} then created a continuous 0.96--12 $\mu$m spectrum by merging the NIRSpec and the MIRI spectrum between 5 and 5.3 $\mu$m, where the spectra overlapped.  The 0.96--12 $\mu$m spectrum is shown in Figure \ref{fig:spectrum} in units of $f_\lambda$ along with the locations of prominent molecular absorption bands of H$_2$O, CH$_4$, CO, CO$_2$, and NH$_3$ identified by \citeauthor{Beiler_2023}

\section{The Method} \label{sec:method}

We use the Brewster retrieval framework \citep{Burningham_2017} for our analysis.  We assume that each datum in the spectrum is generated from the following probabilistic model,

\begin{equation}
F_\lambda(\lambda_i) = (R/d)^2 [\boldsymbol{I}(\lambda_i) * \boldsymbol{\mathcal{F}}_\lambda(\boldsymbol \theta_\textrm{atm}, \lambda_j) ] + \epsilon(\lambda_i), 
\end{equation}
\noindent

\noindent where $F_\lambda(\lambda_i)$ is a random variable giving the flux density of the  spectrum  at the $i$th wavelength $\lambda_i$, $R$ is the radius of the brown dwarf, $d$ is the distance of the brown dwarf, $\boldsymbol{I}(\lambda_i)$ is the instrument profile at $\lambda_i$, the asterisk denotes a convolution, $\boldsymbol{\mathcal{F}}_\lambda$ is a model emergent flux density at the surface of the brown dwarf, $\boldsymbol \theta_\textrm{atm}$ is a vector of parameters describing the atmospheric model, $\lambda_j$ is equal to $\lambda_k + \Delta \lambda$, where $\lambda_k$ is the wavelength at which the model emergent flux is calculated and $\Delta \lambda$ is a parameter that accounts for any wavelength uncertainty, and $\epsilon (\lambda_i)$ is a random variable that is distributed as a Gaussian with a mean of zero and a variance of $\sigma(\lambda_i)^2$.  We further assume the variances $\sigma^2(\lambda_i)$ are given by,
\begin{equation}
\sigma^2(\lambda_i) = s^2(\lambda_i) + 10^b,
\end{equation}
\noindent  where $s_i(\lambda_i)$ is the standard error of the spectrum at $\lambda_i$ and $b$ is a tolerance parameter that is used to inflate the measured uncertainties to account for unaccounted sources of uncertainty \citep[e.g.,][]{Hogg_2010, Foreman_2013, Burningham_2017}.

The one-dimensional atmospheric model is divided into 64 layers (65 levels), with the pressure ranging from 10$^{-4}$ to 10$^{2.3}$ bar, in steps of 0.1 dex. This range was chosen based on the pressure regions that can be probed with the spectrum being used for this retrieval analysis and the available opacities. For simplicity we assume the atmosphere is cloudless and so the only sources of opacity are the absorbing gases H$_2$, He, H$_{2}$O, CH$_{4}$, CO, CO$_{2}$, NH$_{3}$, H$_{2}$S, K, Na, and PH$_{3}$.  H$_2$ and He contribute a continuum opacity in the form of collision-induced absorption (i.e. H$_2$-H$_2$, H$_2$-CH$_4$, and H$_2$-He).  The uniform-with-altitude volume number mixing ratios\footnote{The volume number mixing ratio of a species is the number density of that species divided by the total number density of the gas.} (hereafter mixing ratios) of the remaining molecules are free parameters.  The thermal profile is modeled with a 5-knot interpolating spline in which the knots are located at the top ($T_\textrm{Knot 1}$), middle ($T_\textrm{Knot 3}$) and bottom ($T_\textrm{Knot 5}$) of the atmosphere, with one point halfway between the top and the middle ($T_\textrm{Knot 2}$) of the atmosphere, and one point halfway between the bottom and the middle ($T_\textrm{Knot 4}$) of the atmosphere. The mass and radius of the brown dwarf are also free parameters which are then used calculate the surface gravity ($g=GM/R^2$).  Taken together, the parameters for the mixing ratios of the 9 gas species, the 5 parameters for the thermal profile,  and mass and radius make up $\boldsymbol \theta_\textrm{atm}$ in Equation 1.  

For a given $\boldsymbol \theta_\textrm{atm}$, the emergent spectrum at the top of the atmospheric $\boldsymbol{\mathcal{F}}_\lambda$ is calculated by using a two-stream source function technique from \citet{Toon_1989}. The emergent spectrum is then convolved with the instrument profile $\boldsymbol{I}(\lambda_i)$, which we assume is a Gaussian, to account for the variable resolving power of the data (see \citet{Beiler_2023} for further discussion on this latter process).  

If we let, $\boldsymbol \Theta = \{ \boldsymbol \theta_\textrm{atm}, d, b, \Delta \lambda\}$ then we can use Bayes' Theorem to calculate the posterior probability density function for the parameters $\boldsymbol \Theta$  given the data $\boldsymbol f_\lambda$, 

\begin{equation}
p(\boldsymbol{\Theta}|\boldsymbol{f}_\lambda) = \frac{p(\boldsymbol{\Theta})\mathcal{L}(\boldsymbol{f}_\lambda|\boldsymbol{\Theta})}{p(\boldsymbol{f}_\lambda)},
\end{equation}

\noindent  where $p(\boldsymbol{\Theta})$ is the prior probability for the set of parameters, $\mathcal{L}(\boldsymbol{f}_\lambda|\boldsymbol{\Theta})$ is the likelihood that quantifies the probability of the data given the model, and $p(\boldsymbol{f}_\lambda)$ is the Bayesian evidence.  If we let

\begin{equation}
\mathcal{M}_\lambda (\lambda_i)= (R/d)^2 [\boldsymbol{I}(\lambda_i) * \boldsymbol{\mathcal{F}}_\lambda(\boldsymbol \theta_\textrm{atm}, \lambda_j) ],
\end{equation}
\noindent
then the natural logarithm of the likelihood function is given by,

\begin{equation}
\ln \, \mathcal{L}(\boldsymbol{f}_\lambda|\boldsymbol{\Theta}) =  -\frac{1}{2}\sum^n_{i=1} \left\{  \frac{[f_{\lambda, i} - \mathcal{M}_\lambda (\lambda_i) ]^2}{\sigma(\lambda_i)^2} - \ln[2\pi \sigma(\lambda_i)^2] \right \},
\end{equation}

\noindent
because we assume that the data are independent and $\epsilon(\lambda)$ is distributed as a Gaussian.  The prior distributions for each of the 19 parameters are given in Table \ref{table:priors}. 

\begin{deluxetable*}{cc}[t!]
\centering
\tablecaption{Parameters Priors \label{table:priors}}
\tablehead{
\colhead{Parameter} &
\colhead{Prior\tablenotemark{a}}}
\startdata
    Gas Volume Mixing Ratio $\log (f_i)$\tablenotemark{b,c}  & $\mathcal{U}(-12,\infty), \sum_{i=1}^9 f_i \leq 1$  
    \\
    Mass $M$ ($\mathcal{M}^\textrm{N}_\textrm{Jup}$) & $\mathcal{U}(1, 80)$
    \\
    Radius $R$ ($\mathcal{R}^\mathrm{N}_{e\mathrm{J}}$) & $\mathcal{U}(0.5, 2)$ 
    \\
    Wavelength Shift $\Delta \lambda$ ($\mu$m) & $\mathcal{U}(-0.01, 0.01)$ 
    \\
    Tolerance Factor $b$ &  $\mathcal{U}(10^{0.01}\times \textrm{min}(\sigma_{i}^{2})), 10^{100} \times \textrm{max}(\sigma_{i}^{2}))$ 
    \\
    5-Knot Thermal Profile: $T_{\textrm{Knot i}} (\textrm{K})$  & $\mathcal{U}(0, 5000)$ 
    \\
    Madhusudhan \& Seager Thermal Profile: $\alpha_{1}$, $\alpha_{2}$, $P_1$, $P_3$, $T_3$ & $\mathcal{U}(0.25, 0.5)$, $\mathcal{U}(0.1, 0.2)$, $\mathcal{U}(10^{-4}, 10^{2.3})$, $\mathcal{U}(10^{-4}, 10^{2.3})$, $\mathcal{U}(0, 5000)$
    \\
    Distance $d$ (pc) & $\mathcal{N}(13.57, 0.37^{2})$
\enddata

\tablenotetext{a}{$\mathcal{U}(\alpha, \beta)$ denotes a uniform distribution between $\alpha$ and $\beta$ while $\mathcal{N}(\mu, \sigma^2)$ denotes a normal distribution with a mean of $\mu$ and a variance of $\sigma^2$.}
\tablenotetext{b}{We included H$_{2}$O, CH$_{4}$, CO, CO$_{2}$, NH$_{3}$, H$_{2}$S, K, Na, PH$_{3}$.}
\tablenotetext{c}{All volume mixing ratios are reported as the log of the volume number mixing ratio (the number density of the species divided by the total number number density of the gas), where the remainder of the gas is assumed to be H$_{2}$-He (1$-$$\sum_i f_i$). Of the remainder gas 84\% of the volume mixing ratio is from H$_{2}$ and 16\% is from He, assuming a solar abundance of 91.2\% of number of atoms of H and 8.7\% of number of atoms of He \citep{Asplund_2009}.}
\end{deluxetable*}

To explore the posterior parameter space, we use the nested sampling version of the Brewster, which uses PyMultiNest \citep{Buchner_2014}. PyMultiNest is initialized to sample the parameter space with 500 live points for 19 free parameters. The calculation was done using the Owens cluster \citep{Owens_2016} at the Ohio Supercomputer Center \citep{OSC_1987}.  The sampling is complete when the change in the natural logarithm of the evidence is less than 0.5 \citep[for a deeper discussion see][]{Feroz_2009, Speagle_2020}.

\section{Results \& Discussion} \label{sec:results}

The result of solving Bayes' theorem is a joint posterior distribution for the 19 parameters.  In the Appendix, Figure \ref{fig:Corner_5_knot_complete} shows the marginalized posterior probability distributions for all 19 parameters using equally weighted posterior samples generated by PyMultiNest and Table \ref{table:4} gives the median, and 1$\sigma$ uncertainty for each of the parameters.  In the following sections, we discuss the values of these parameters in more detail.

\subsection{The Thermal Profile}
\label{sec:5knotprofile}

\begin{figure*}[htb!]
    \centering
    \includegraphics[width=\linewidth]{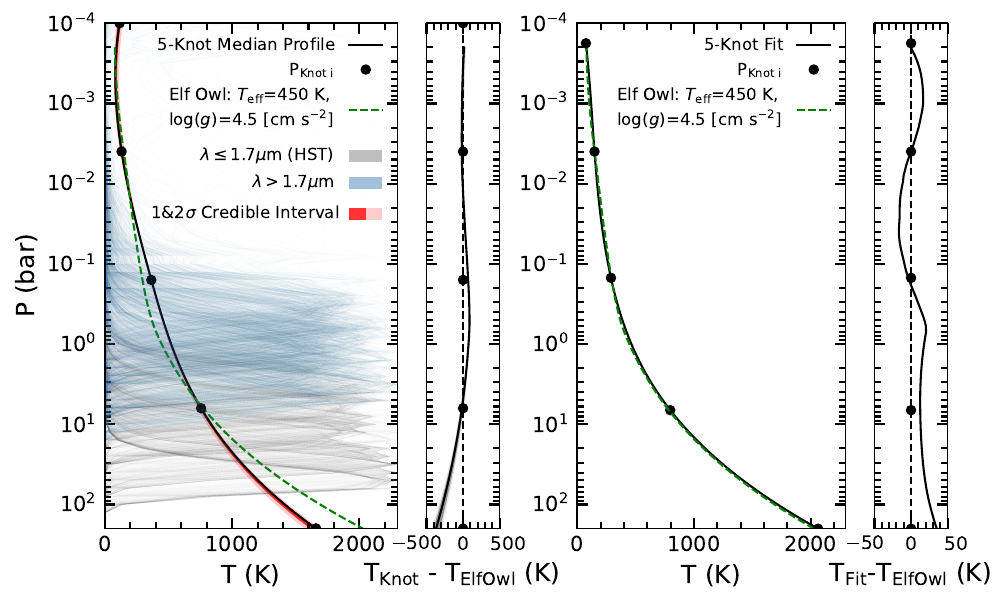}
    \caption{\textbf{Left two panels (Retrieved vs. Forward Model):} the black solid curve shows the 5-knot retrieved median thermal profile, the red region shows the 1 and 2$\sigma$ central credible interval around the median profile, and the green dashed curve is a solar metallicity, solar C/O ratio, cloudless Elf-Owl thermal profile with $T_\textrm{eff}$=450 K and log($g$)=4.5 [cm s$^{-2}$].  Also plotted are normalized contribution functions in gray (HST wavelength coverage, $\lambda \leq 1.7\, \mu$m) and in blue ($\lambda > 1.7\, \mu$m).  The opacity windows centered at the $J$- and $H$-bands probe deep, hotter layers of the atmosphere while longer wavelengths general probe higher and cooler layers of the atmosphere.  The handful of contributions functions at pressure lower than $10^{-2}$ bar come primarily from the 6.3--7.8 $\mu$m wavelength range. \textbf{Right two panels (Forward Model vs. Forward Model Fit):} the dashed green curve is the Elf-Owl thermal profile with a $T_\textrm{eff}$=450 K and log($g$)=4.5 cm s$^{-2}$ and the black solid curve shows the 5-knot spline fit of the same Elf-Owl thermal profile. The black dots in all the panels represent the position of the 5 knots. \textbf{Note:} The temperature range in panel 2 is an order of magnitude larger than the range in panel 4.}
   \label{fig:T_P_compare_2}
\end{figure*}

The first panel in Figure \ref{fig:T_P_compare_2} shows the retrieved thermal profile; the black solid line shows the median (50th percentile) profile (calculated using the median values of the retrieved parameters) and the red shaded region represents the 16th and 84th percentile ($1\sigma$ central credible interval\footnote{A Bayesian central credible interval gives the range of values in a parameter's posterior distribution that contain $\alpha$\% of the probability.  In contrast, a frequentist $\alpha$\% confidence interval means that $\alpha$\% of a large number of confidence intervals computed in the same way would contain the true value of the parameter.}), and 2.4th and 97.6th percentile ($2\sigma$ central credible interval).   Also plotted are a subset of normalized contribution functions at wavelengths covering the 0.96--12 $\mu$m wavelength range; those in grey are the wavelengths covered by HST/WFC3 spectra ($\lambda < 1.7 \mu$m) while the blue cover the wavelengths longward of $1.7 \mu$m.  Integration of a contribution function over (log) pressure in a semi-infinite atmosphere gives the specific intensity at the top of the atmosphere at the corresponding wavelength \citep{Chamberlain_1987}.  A normalized contribution function therefore indicates the layers of the atmospheres from which light at that wavelength emerges.  The opacity windows centered at the $J$- and $H$-bands probe deep, hotter layers of the atmosphere (grey lines) while longer wavelengths general probe higher and cooler layers of the atmosphere (blue lines).  The handful of contributions functions at pressure lower than $10^{-2}$ bar come primarily from the 6.3--7.8 $\mu$m wavelength range.  The JWST spectrum therefore probes nearly four orders of magnitude in pressure; two more than previous work \citep{Zalesky_2019} using HST spectra alone.   In addition, the median width of the $1\sigma$ central credible interval is $\sim$20 K which is an order of magnitude lower than typically found using HST spectra alone \citep{Zalesky_2019}.  

A cloudless self-consistent 1D radiative-convective equilibrium Sonora Elf Owl thermal profile with solar a metallicity and C/O ratio \citep[green dashed line,][]{Mukherjee_2024_sonora} is also plotted in the first panel of Figure \ref{fig:T_P_compare_2}.  The effective temperature and (log) surface gravity of 450 K and 4.5 [cm s$^{-2}$] were chosen to match our derived values of 458 K and 4.46 [cm s$^{-2}$] (see \S \ref{sec:physicalproperties}) as closely as possible.   The difference between the two profiles are shown in the second panel of Figure \ref{fig:T_P_compare_2}.  Overall, the retrieved profile matches the self-consistent profile well, although the retrieved profile is systematically hotter by up to 100 K between 0.01 and 10 bars and systematically cooler by up to 500 K in the deepest layers of the atmosphere. The retrieved profile also shows a slight temperature reversal of $\sim$30 K at the top of the atmosphere.  While this is probably unphysical, \citet{Faherty_2024} did identify CH$_4$ emission in the moderate-resolution JWST spectrum of the Y dwarf CWISEP J193518.59$-$154620.3 at 3.326 $\mu$m.  They modelled this as a 300 K temperature reversal between the 1 and 10 millibar pressure range and so further investigation into our reversal is warranted.

In order to investigate the possibility that the differences between the retrieved profile and Elf Owl profile are due to an inability of the 5-knot spline to reproduce the shape of the Elf Owl profile, we have fitted the Elf Owl profile with a 5-knot spline and the results are shown in the third panel of Figure \ref{fig:T_P_compare_2}; the difference between the two profiles is shown in the last panel.  The Elf Owl profile does not extend up to the $10^{-4}$ bar level so we placed the top knot at $10^{-3.7}$ bar, the vertical extent of the Elf Owl profile.  The 5-knot spline easily reproduces the Elf Owl profile with a root mean squared deviation of 17 K and a maximum deviation of $<50$ K.  This indicates that the differences between the retrieved profile and the Elf Owl profile are real and statistically signficant.  
 
\subsection{Retrieved Model Spectrum}

\begin{figure*}[htb!]
    \centering
    \includegraphics[width=\linewidth]{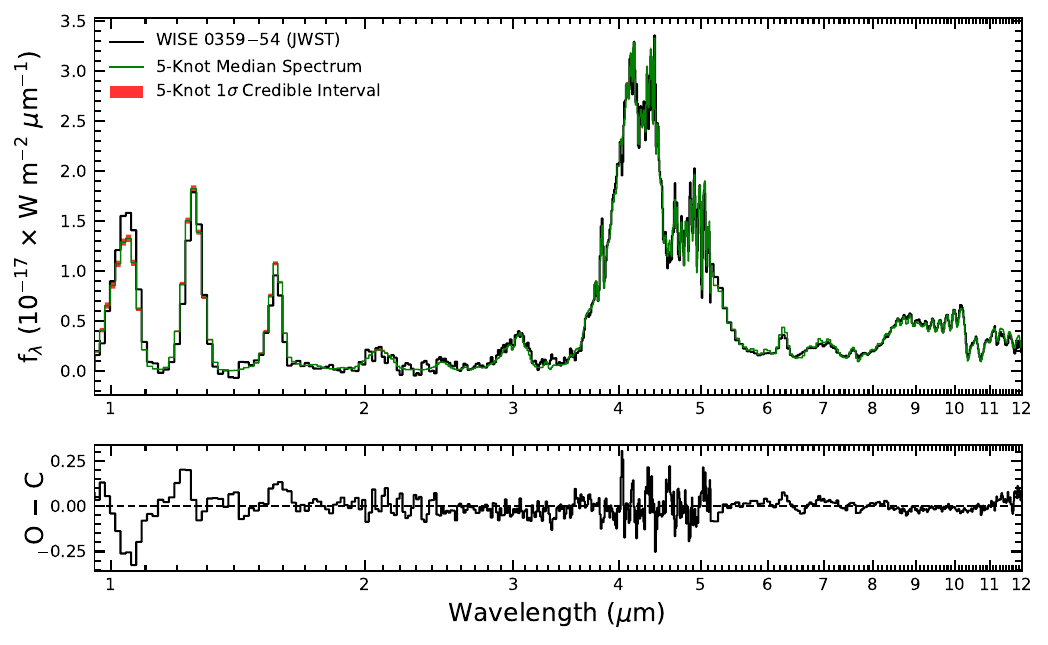}
    \caption{\textbf{Top panel:} shows the observed spectrum in black and the retrieved median spectrum from 5-knot retrieval in green for WISE 0359--54 spectrum covering 0.96-12$\mu$m. The red region show the 1$\sigma$ central credible interval around the median spectrum. \textbf{Bottom panel:} shows the residual spectrum calculated by taking the difference between the retrieved median spectrum and the observed spectrum.}
   \label{fig:spec_free}
\end{figure*}

Figure \ref{fig:spec_free} shows the JWST spectrum of WISE 0359$-$54 along with the retrieved median model spectrum (upper panel) and the residual (O-C, lower panel). The model spectrum is generated using the median thermal profile and the 1$\sigma$ central credible interval is generated using the 1$\sigma$ of the 5-knot thermal profile.  Overall, the model fits the data well as the residuals are mostly random.  However, the model fails to reproduce the observations in the 1--2 $\mu$m range.  The poor agreement shortward of 1.1 $\mu$m is likely a result of our poor understanding of the exact shape of the pressure-broadened wings of the resonant \ion{K}{1} and \ion{Na}{1} doublets at 7665/7699 \AA\, and 8183/8195 \AA, respectively (see \citet{Burningham_2017} for a more in-depth discussion).

\subsection{Mixing Ratios} \label{mixingratios}

\begin{figure*}[htb!]
    \centering
    \includegraphics[width=\linewidth]{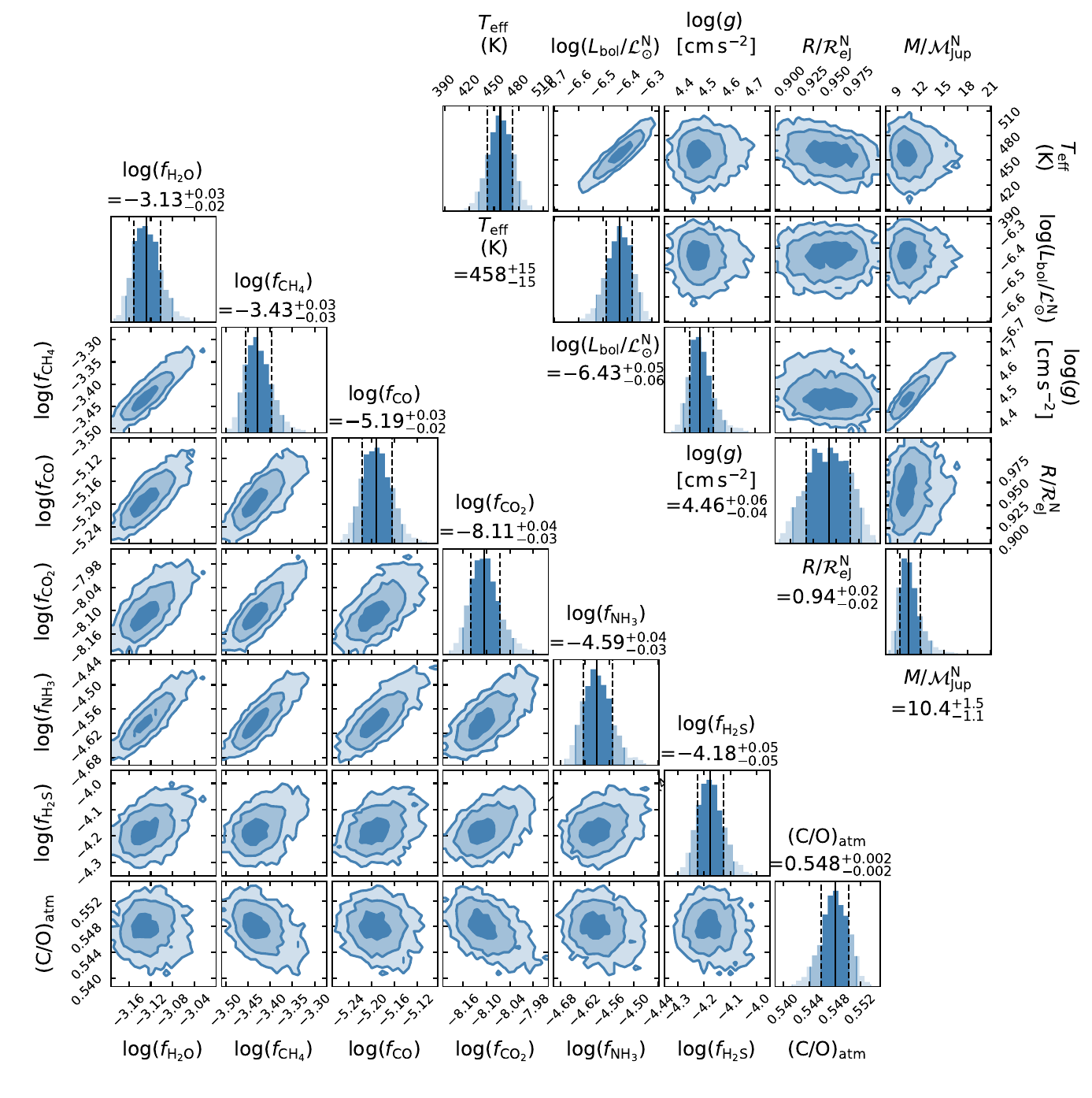}
    \caption{\textbf{Lower left:} Marginalized posterior probability distributions for H$_{2}$O, CH$_{4}$, CO, CO$_{2}$, NH$_{3}$, H$_{2}$S, and atmospheric (C/O)$_\textrm{atm}$ from the 5-knot retrieval for WISE 0359--54. \textbf{Upper right:} Marginalized posterior probability distributions for $T_\textrm{eff}$, $L_\textrm{bol}$, $\log (g)$, $\mathcal{R}^\mathrm{N}_{e\mathrm{J}}$ and $\mathcal{M}^\textrm{N}_\textrm{Jup}$ from the 5-knot retrieval for WISE 0359--54. In both panels, the values above the histograms represent the parametric median (50th percentile) values with the errors representing the $1\sigma$ (16th and 84th percentile) values.  The different shades in the 1D and 2D histograms represent the 1, 2 and $3\sigma$ central credible interval, respectively, with the darkest shade corresponding to $1\sigma$.}
   \label{fig:Corner_free}
\end{figure*}

\begin{figure}[htb!]
    \centering
    \includegraphics[width=\linewidth]{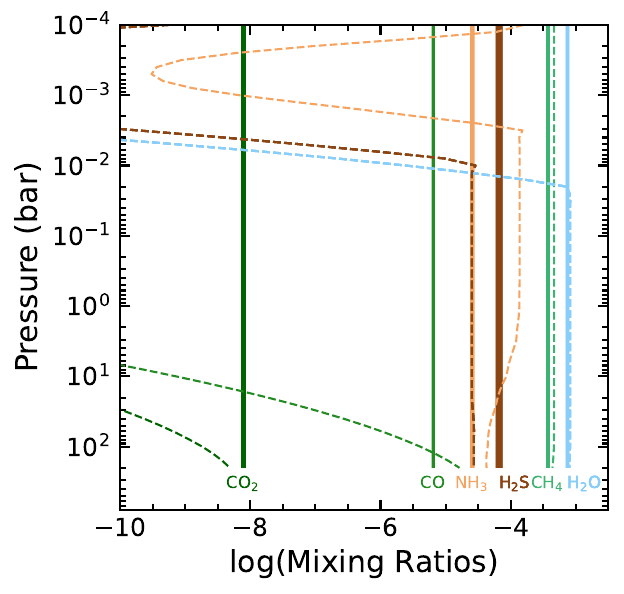}
    \caption{Retrieved uniform-with-altitude median mixing ratios with 1$\sigma$ central credible interval (shaded) for H$_2$O, CH$_4$, CO, CO$_2$, NH$_3$, and H$_2$S for WISE 0359--54 from the 5-knot retrieval.  Also shown are the predicted thermochemical equilibrium mixing ratios (dashed).}
   \label{fig:vmr_free}
\end{figure}

Figure \ref{fig:Corner_free} shows the marginalized posterior probability distributions for the mixing ratios of H$_2$O, CH$_4$, CO, CO$_2$, NH$_3$, and H$_2$S.  With secure detections of all the dominant carbon- and oxygen-bearing molecules,  we can also calculate the atmospheric (C/O)$_\textrm{atm}$ ratio as

\begin{equation}
(\textrm{C}/\textrm{O})_\textrm{atm} = \frac{f_{\textrm{CO}} + f_{\textrm{CO}_{2}} + f_{\textrm{CH}_{4}}}{f_{\textrm{H}_{2}\textrm{O}} + f_{\textrm{CO}} + 2f_{\textrm{CO}_{2}}}.  
\end{equation}

\noindent
and so Figure \ref{fig:Corner_free} also shows the marginalized posterior probability distribution for (C/O)$_\textrm{atm}$ calculated using the samples of $f_\textrm{CO}$, $f_{\textrm{CO}_2}$, $f_{\textrm{CH}_4}$, and $f_{\textrm{H}_2\textrm{O}}$. It should be noted that $\sim $20\% of oxygen is depleted due to the sequestration of oxygen in condensates like enstatite (MgSiO$_3$) and forsterite (MgSi$_2$O$_4$) \citep{Lodders_2002} which would bring the median bulk $(\textrm{C}/\textrm{O})_\textrm{bulk}$ to 0.658.

In order to perform a sanity check on our retrieved abundances and (C/O)$_\textrm{atm}$ ratio, we compare our values to those reported by  \citet{Zalesky_2019} for 8 Y dwarfs, and \citet{Barrado_2023} and \citet{Lew_2024} for the archetype Y dwarf WISE 1828$+$26 in Table \ref{tab:ratiocomps}. We note that the \citeauthor{Barrado_2023} uncertainties are an order-of-magnitude larger than the other works because they were computed by combining (with equal weight) the posterior distributions from five different retrieval analyses.  

In general, the mixing ratio and (C/O)$_\textrm{atm}$ values agree well.  The mixing ratios of H$_2$O and NH$_3$ fall within the range of values found by  \citet{Zalesky_2019} but the values for CH$_4$ and (C/O)$_\textrm{atm}$ fall towards the lower and upper limits of the ranges, respectively.   Our values and those of \citet{Lew_2024} are inconsistent given the uncertainties; however this could be because WISE 1828$+$26 is 110 K cooler than WISE 0359$-$54 and/or because both sets of measurements are likely dominated by systematic uncertainties not accounted for in the respective analyses (see \S  \ref{sec:MadhusudhanProfile}).  The \citeauthor{Barrado_2023} values generally agree with our values, but this is more likely a result of their order-of-magnitude-larger uncertainties generated by combining the results of several retrieval analyses.  

H$_{2}$S exhibits many rotation-vibrational bands in the 1--12 $\mu$m wavelength range centered at 1.33, 1.6, 2, 2.6, 4.0, and 8.0  $\mu$m.  However, with the exception of a single absorption line detected at $\lambda$=1.590 $\mu$m in an $R\approx 45,000$ spectrum of the T6 dwarf 2MASS J08173001$-$6155158 \citep{Tannock_2022}, spectral features of H$_2$S have remained undetected in the spectra of cool brown dwarfs.  However, \citet{Hood_2023} showed that a retrieval that includes H$_2$S as an opacity source 
produced a better fit to the moderate-resolution ($R\sim 6000$) near-infrared spectrum of the T9 dwarf UGPS  J072227.51$-$054031.2 than a retrieval without H$_2$S opacity.  \citeauthor{Lew_2024} also found that excluding H$_2$S opacity in their retrieval of WISE 1828$+$26 increased the $\chi^2$ of the fit by over 900.  These results suggest that retrievals can still detect H$_2$S in the atmospheres of cool brown dwarfs even though there are no obvious absorption features in their low- to moderate-resolution spectra.  \citeauthor{Lew_2024} retrieved a mixing ratio of $-4.44^{+0.03}_{-0.03}$ for WISE 1828$+$26, which is 0.24 dex lower than an our value.  We note that these are the only two detections of H$_2$S in atmospheres of Y dwarfs and so a larger sample of cool brown dwarfs will be required (Kothari et al., in prep) to determine whether this difference is significant or not.

Finally, we included PH$_3$ as a source of opacity in our retrieval because the best fitting Sonora model for WISE 0359$-$54 in \citet{Beiler_2023} predicts the presence of phosphine.  However, our retrieved mixing ratio of $-10.00^{+1.15}_{-1.26}$ is consistent with the lack of any PH$_3$ spectroscopic features \citep{Beiler_2023}.   The lack of PH$_3$ absorption bands in the spectra of the coolest brown dwarfs (down to $\sim$250 K) \citep{Miles_2020, Luhman_2023}  remains an outstanding problem given that PH$_3$ has been detected in the spectra of Jupiter and Saturn \citep{Gillett_1973, Beer_1975, Bregman_1975, Barshay_1978}.

\begin{deluxetable*}{l c c c c c}[htb!]
    \tablecolumns{6}
    \centering
    \tablecaption{\label{tab:ratiocomps}Parametric Value Comparison}
    \tablehead{
    \multicolumn{1}{c}{Parameter} &
    \multicolumn{2}{c}{W0359--54} &
    \multicolumn{1}{c}{8 Y dwarfs} &
    \multicolumn{2}{c}{WISE J1828$+$26}
    \\
    \cline{5-6}
    \multicolumn{1}{c}{} &
    \multicolumn{2}{c}{This Work} &
    \multicolumn{1}{c}{\citet{Zalesky_2019}} &
    \multicolumn{1}{c}{\citet{Lew_2024}} &
    \multicolumn{1}{c}{\citet{Barrado_2023}\tablenotemark{a}}
    \\
    \cline{2-3}
    \multicolumn{1}{c}{} &
    \multicolumn{1}{c}{5-Knot} &
    \multicolumn{1}{c}{Parametrized} &
    \multicolumn{1}{c}{(min--max)} &
    \multicolumn{1}{c}{} &
    \multicolumn{1}{c}{}
    } 
    \startdata
    $\log (f_\textrm{H$_{2}$O})$ &  $-$3.13$^{+0.03}_{-0.02}$ &  $-3.10^{+0.04}_{-0.04}$  &  $-$2.68 -- $-$3.32 &  $-2.71^{+0.01}_{-0.02}$ &  $-3.03^{+0.18}_{-0.21}$ \\ 
    $\log (f_\textrm{CH$_{4}$})$ & $-3.43^{+0.03}_{-0.03}$ &   $-3.34^{+0.04}_{-0.04}$  &  $-$2.63 -- $-$3.42 &  $-3.07^{+0.01}_{-0.02}$ &   $-3.65^{+0.11}_{-0.21}$ \\
    $\log (f_\textrm{CO})$ &  $-5.19^{+0.03}_{-0.02}$ &  $-5.18^{+0.04}_{-0.04}$ & \phantom{-,} $-$3.3 -- $-$4.2 \tablenotemark{b} & $\cdots$ & $\cdots$ \\
    $\log (f_\textrm{CO$_{2}$})$ &  $-8.11^{+0.04}_{-0.03}$ &  $-8.05^{+0.05}_{-0.05}$ & \phantom{-,} $-$3.6 -- $-$4.6 \tablenotemark{b} &  $-8.79^{+0.03}_{-0.04}$  & $\cdots$ \\
    $\log(f_\textrm{NH$_{3}$})$ &  $-4.59^{+0.04}_{-0.03}$ &  $-4.51^{+0.05}_{-0.05}$ &  $-$4.11 -- $-$4.84 &  $-4.21^{+0.02}_{-0.02}$ &   $-4.79^{+0.15}_{-0.25}$ \\ 
    $\log(f_\textrm{H$_{2}$S})$ &  $-4.18^{+0.05}_{-0.05}$ &  $-4.60^{+0.12}_{-0.12}$ & \phantom{-,} $-$4.3 -- $-$6.3 \tablenotemark{b} &  $-4.44^{+0.03}_{-0.03}$  & $\cdots$ \\ 
    C/O & \phantom{$-$0} $0.548^{+0.002}_{-0.002}$ & \phantom{$-$0} $0.538^{+0.003}_{-0.002}$ &  $-$0.55 -- $-$1.10 & \phantom{0} $0.45^{+0.01}_{-0.01}$ & \phantom{0} $0.21^{+0.45}_{-0.03}$ \\
    \enddata
    
    \tablecomments{\textsuperscript{a}{Averaged retrieved results from 5 different retrieval codes.} \\
    \textsuperscript{b}{These ranges represent 3$\sigma$ upper limit values.} 
    }
    
\end{deluxetable*}

Figure \ref{fig:vmr_free} shows a comparison of the mixing ratios for H$_2$O, CH$_4$, CO, CO$_2$, NH$_3$, and H$_2$S to the predictions of a thermochemical equilibrium model.  The solid colored bars indicate the 1$\sigma$ central credible interval for each mixing ratio and the corresponding dashed line gives the model predictions which are calculated using chemical equilibrium grids generated using the NASA Gibbs minimization CEA code \citep[see][]{Fegley_1994, Lodders_1996, Lodders_1999_single, Lodders_2002_single, Lodders_2010_single, Lodders_2002, Lodders_2006, Visscher_2006, Visscher_2010, Visscher_2012, Moses_2013} at solar metallicity and C/O.  The retrieved values are uniform-with-altitude and so show no variation with pressure, while the model predictions are calculated along the retrieved thermal profile (see \S\ref{sec:5knotprofile}) and so do show variations with pressure.  The rapid decrease in the model mixing ratios of H$_2$O, H$_2$S , and NH$_3$ above $\sim 10^{-2}$ bar are a result of these species condensing out of the gas phase into water ice, ammonia ice, and NH$_4$SH (solid).    The rapid increase in the mixing ratio of NH$_3$ above $10^{-3.2}$ bar is a result of the slight temperature reversal at the top of the thermal profile that is likely not physical (see \S\ref{sec:5knotprofile}).    

The mixing ratio values of both H$_2$O and CH$_4$ indicate they are the most abundant species in the atmosphere and they agree well with the predictions.  The NH$_3$ mixing ratio is 0.6 dex lower than the model predicts at the nominal pressure of 1 bar, while the mixing ratios of CO and CO$_2$ are orders of magnitude higher at 1 bar.  All three of these mismatches can be ascribed to disequilibrium chemistry due to vertical mixing in the atmosphere \citep{Lodders_1996, Saumon_2000, Hubeny_2007}.  We defer a discussion of this disequilibrium chemistry to \S \ref{sec:Kzz} where we attempt to measure the vigor of this mixing using the retrieved mixing ratios and a 1D chemical kinetics forward modeling framework.  Finally, the mixing ratio of H$_2$S is 0.4 dex (2.5$\times$) higher than the model predicts.

\subsection{Physical properties: $M$, $R$, $L_\textrm{bol}$, $g$, and $T_\textrm{eff}$}
\label{sec:physicalproperties}

The marginalized posterior distributions for $M$ and $R$ are shown in Figure \ref{fig:Corner_free}.  The $M$ and $R$ posterior samples can be used to calculate the posterior distribution for surface gravity ($g = MG/R^2$) and so the distribution of $\log (g)$ [cm s$^{-2}$] is also shown in Figure \ref{fig:Corner_free}.  The bolometric flux $F_\textrm{bol}$ distribution can be calculated by integrating model spectra over all wavelengths.  To account for light emerging at wavelengths shorter than 0.96 $\mu$m and longer than 12.0 $\mu$m, we linearly interpolated the model from 0.96 $\mu$m to zero flux at zero wavelength and then extended the model to $\lambda=\infty$ using a Rayleigh-Jeans tail where $f_{\lambda,\textrm{RJ}} \propto \lambda^{-4}$; the constant of proportionality is calculated  using the flux density of the last model wavelength.  The bolometric luminosity is then given by $L_\mathrm{bol} = 4 \pi d^2 F_\mathrm{bol}$, where $d$ is the retrieved distance to the object, which results in the posterior distribution of $\log (L_\textrm{bol}/\mathcal{L}^\textrm{N}_\odot)$ shown in Figure \ref{fig:Corner_free}.  Finally, we compute the effective temperature distribution shown in Figure \ref{fig:Corner_free} using the $L_\textrm{bol}$ and $R$ values and the Stefan-Boltzman Law, 

\begin{equation}
T_\mathrm{eff} = \left ( {\frac{L_\mathrm{bol}}{4 \pi \sigma R^2}}\right) ^{\frac{1}{4}}.
\end{equation}

The retrieved mass of WISE 0359$-$54 is 10.4$^{+1.5}_{-1.1}$ $\mathcal{M}^\textrm{N}_\textrm{Jup}$, where $ \mathcal{M}^\textrm{N}_\textrm{Jup}$ is the nominal Jupiter mass \citep[assuming $G = 6.67430 \times 10^{-11} \textrm{ m}^3 \textrm{ kg}^{-1} \textrm{ s}^{-2}$,][]{Mamajek_2015}.   This value falls at the lower end of the $\sim$9--31$\mathcal{M}^\textrm{N}_\textrm{Jup}$ range reported in \citet{Beiler_2023} who used the observed bolometric luminosity of WISE 0359$-$54, an assumed age range of 1--10 Gyr, and the Sonora Bobcat solar metallicity evolutionary models \citep{Marley_2021} to estimate the mass of WISE 0359$-$54. 

The retrieved radius is found to be 0.94$\pm$0.02 $\mathcal{R}^\mathrm{N}_{e\mathrm{J}}$, where $ \mathcal{R}^\mathrm{N}_{e\mathrm{J}}$ is Jupiter’s nominal equatorial radius of 7.1492 $\times$ 10$^{7}$ m \citep{Mamajek_2015}. This is consistent with the value reported by \citet{Beiler_2023} who used the observed bolometric luminosity of WISE 0359$-$54, an assumed age range of 1--10 Gyr, and the Sonora Bobcat solar metallicity evolutionary models \citep{Marley_2021} to find 0.94$^{+0.074}_{-0.057}$  $\mathcal{R}^\mathrm{N}_{e\mathrm{J}}$ from a Monte Carlo simulation.

\begin{figure}[htb!]
    \centering
    \includegraphics[width=\linewidth]{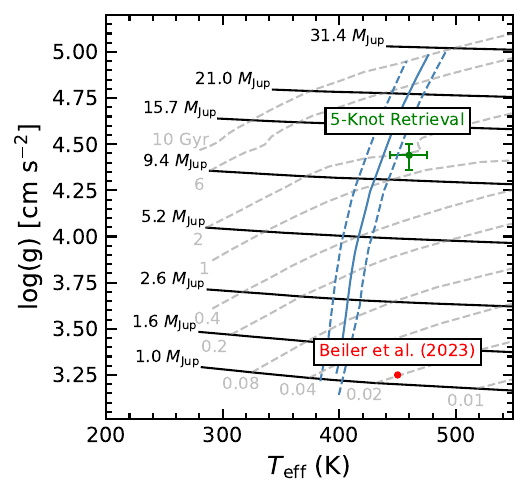}
    \caption{Evolution of Bobcat Sonora solar metallicity cloudless brown dwarfs in the effective temperature surface gravity plane \citep{Marley_2021}. The black lines are cooling tracks for brown dwarfs with masses of 31.4, 21, 15.7, 9.4, 5.2, 2.6, 1.6 and 1 $\mathcal{M}^\textrm{N}_\textrm{Jup}$, while the gray lines are isochrones for ages of 10, 6, 2, 1, 0.4, 0.2, 0.08, 0.04, 0.02, and 0.01 Gyr. The loci of points with bolometric luminosities equal to that of WISE 0359--54 for ages between 0.1 and 10 Gyr are shown as the solid near-vertical lines while the $\pm$1$\sigma$ uncertainties on the bolometric luminosities are shown as dotted lines. The blue dot shows the $T_\textrm{eff}$ and log($g$) value calculated using the 5-knot retrieved results with the horizontal and vertical error bar representing 1$\sigma$ interval for $T_\textrm{eff}$ and log($g$), respectively. The red dot is the best-fit Sonora Bobcat model in \citet{Beiler_2023}}
   \label{fig:Sams_fig_6_compare}
\end{figure}
 
The bolometric luminosity of $\log  (L_\textrm{bol}/\mathcal{L}^\textrm{N}_\odot)=-6.43^{+0.05}_{-0.06}$ is similar to the value of $\log  (L_\textrm{bol}/\mathcal{L}^\textrm{N}_\odot)=-6.400\pm 0.025$ reported by \citet{Beiler_2023}.


The retrieved surface gravity is $\log g$ = 4.46$^{+0.06}_{-0.04}$ [cm s$^{-2}$] and the retrieved effective temperature is $T_\textrm{eff} = 458^{+15}_{-15}$ K.  Figure \ref{fig:Sams_fig_6_compare} shows cloudless evolutionary models in the effective temperature/surface gravity plane with the position of WISE 0359$-$54 indicated.  The loci of points with bolometric luminosities equal to that of WISE 0359–54 for ages between 0.1 and 10 Gyr is shown as a near-vertical line.  The discrepancy between the two is most likely a result of the fact that the model does not extend to infinite wavelengths and thus our bolometric flux is systematically low.  Also plotted is the best-fit effective temperature and surface gravity from \citet{Beiler_2023} who used a custom grid of Sonora Cholla models \citep{Karalidi_2021}  that includes an additional parameter $K_\textrm{zz}$, the vertical eddy diffusion coefficient.  The \citeauthor{Beiler_2023} surface gravity is uncomfortably low resulting an age estimate of $\sim$20 Myr.   Our retrieved values gives an age of $\sim$2 Gyr which is more consistent with the age estimates of the field population of warmer brown dwarfs \citep{Dupuy_2017, Best_2024}.

\subsection{Constraints on the eddy diffusion parameter-- {\kzz}}
\label{sec:Kzz}

Vertical atmospheric dynamics can significantly alter the photospheric abundance of gases like CH$_4$, NH$_3$, CO, and CO$_2$ by dredging them up from the hotter deeper atmosphere across several pressure scale heights. Exactly how much of the photospheric abundances are disturbed away from thermochemical equilibrium depends on the vigor of vertical mixing in the atmosphere of these objects \citep{Fegley_1994,hubeny07,visscher2011,Zahnle14,Philips20,karilidi2021,Mukherjee22,lacy23,lee23}. The strength of vertical mixing is often quantified using the vertical eddy diffusion parameter-- {\kzz}. The {\kzz} parameter quantifies the rate of overturning motion occuring in the atmosphere and a  higher {\kzz} represents more vigorous vertical mixing. But {\kzz} has remained uncertain (even in the solar system giants) by several orders of magnitude until now mainly because of the lack of access to high SNR spectra of brown dwarfs in the infrared which can facilitate very precise constraints on atmospheric chemical abundances. The very precise constraints on abundances of various gases obtained in this work makes it a perfect target to constrain {\kzz} in its deep atmosphere.

In order to obtain constraints on {\kzz} from our retrieved gas abundances, we use the chemical kinetics model {\it Photochem} \citep{wogan23}. We use the median retrieved 5-knot thermal profile as an input to the chemical kinetics model along with the median $\log(g)$ constraints obtained by our 5-knot retrieval model. Using these inputs, we generate a grid of chemical forward models with {\it Photochem} by varying three key parameters that can influence chemistry of brown dwarfs -- atmospheric metallicity, atmospheric (C/O)$_\textrm{atm}$ ratio, and {\kzz}. For a given (C/O)$_\textrm{atm}$, we remove about 20\% of the O- from gas phase assuming it is used up in condensates in the deeper atmosphere. Our chemical forward model grid samples metallicities from sub-solar to super-solar values between $-$0.3 to $+$0.3 with an increment of 0.1 dex except between $-$0.2 to $+$0.1, for which the increment is even smaller at 0.02 dex. We also vary the (C/O)$_\textrm{atm}$ ratio from sub-solar to super-solar values of 0.5 to 1.5$\times$ (C/O)$_\odot$, where the (C/O)$_\odot$ is assumed to be 0.458. We vary $\log$({\kzz}) from 2 to 11 with an increment of 1 except between the values of 6 to 10 where we include a finer sampling of 0.5. These {\kzz} values are in cm$^2$ s$^{-1}$.

We use the extensive grid of chemical forward models to fit the retrieved abundances  with the model abundance profiles of CH$_4$, CO, CO$_2$, H$_2$O, and NH$_3$ at a pressure of 0.1 bars. We choose this pressure because it is smaller than the minimum quench pressures expected for these gases for the range of {\kzz} used in this work. For each forward model, we define a combined $\chi^2$ using,
\begin{equation}
    \chi^2 = \sum_{X} \left(\dfrac{X_{\rm ret}- X_{\rm model}(0.1 {\rm bar})}{{\sigma}^X_{\rm ret}} \right)^2
\end{equation}

\noindent
where $X_{\rm ret}$ is the retrieved abundance of gas $X$, $X_{\rm model}(0.1 {\rm bar})$ is the abundance of the same gas at 0.1 bars in the forward model grid, and ${\sigma}^X_{\rm ret}$ is the retrieved uncertainty on the abundance of gas $X$. We calculate the $\chi^2$ of all our chemical models using this formulation and then produce a corner-plot for the sampled parameter points in our grid using w=$e^{-\chi^2/2}$ as weight for each sampled grid point. 

\begin{figure}[htb!]
    \centering
    \includegraphics[width=\linewidth]{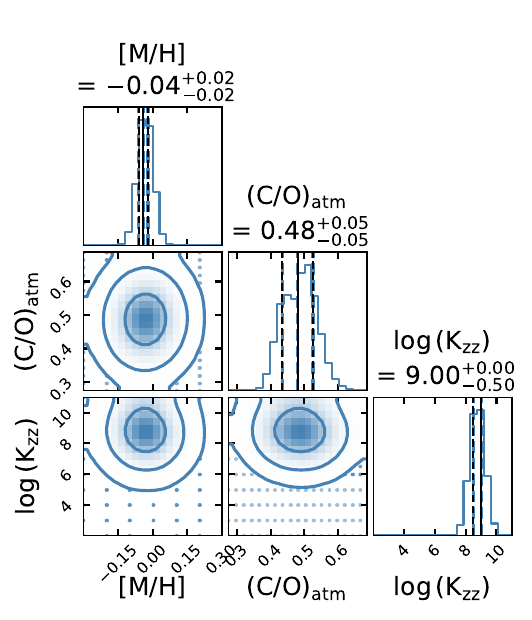}
    \caption{Corner plot showing the constraints on metallicity [M/H], (C/O)$_\textrm{atm}$, and {\kzz} obtained by fitting the retrieved gaseous abundances of CH$_4$, CO, NH$_3$, H$_2$O, and CO$_2$ with a grid of disequilibrium chemistry forward models. The grid of disequilibrium chemistry forward models uses the retrieved 5-knot $T(P)$ profile constraint as an input and calculates the chemical abundance profiles across a large range of metallicity, (C/O)$_\textrm{atm}$, and {\kzz}. Gaseous abundances obtained in the 5-knot retrieval were used for this analysis.}
   \label{fig:corner_photochem}
\end{figure}

\begin{figure*}[htb!]
    \centering
    \includegraphics[width=\linewidth]{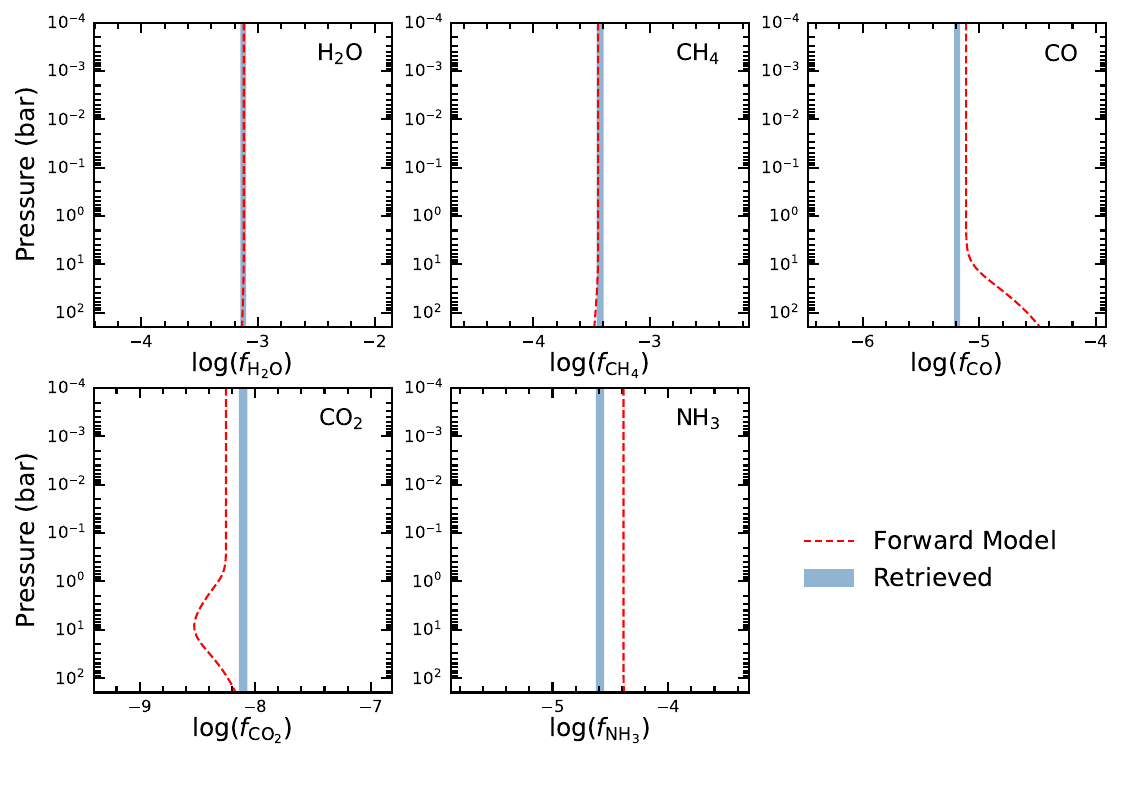}
    \caption{Comparison of the abundances in the best-fit disequilibrium chemistry forward model from {\it Photochem} with the retrieved abundances from the 5-knot thermal profile. Retrieved volume mixing ratios of CH$_4$, CO, NH$_3$, H$_2$O, and CO$_2$ are shown with blue shaded regions (representing 1$\sigma$ central credible interval) in each panel, whereas the red lines show the profiles from the best-fit forward model. }
   \label{fig:abundances_photochem}
\end{figure*}

Figure \ref{fig:corner_photochem} shows this corner plot depicting our constraints on the atmospheric metallicity, (C/O)$_\textrm{atm}$ ratio, and {\kzz} obtained from the $T(P)$ profile and abundances retrieved using the 5-knot modeling setup. The best-fit forward model abundance profiles for CH$_4$, CO, CO$_2$, H$_2$O, and NH$_3$ along with the retrieved abundances are shown in Figure \ref{fig:abundances_photochem}. This analysis finds that the atmospheric metallicity of the object is very slightly sub-solar and the (C/O)$_\textrm{atm}$ ratio is $\sim$ 0.48. We note that 20\% of the O- has been removed out of the gas phase which means that the actual bulk gas phase (C/O)$_\textrm{bulk}$ in the deep atmosphere in this best-fit model is $\sim$0.58. 

The best-fit {\kzz} value is found to be 10$^9$ cm$^2$s$^{-1}$, which is relatively large compared to previous estimates of {\kzz} in the atmospheres of cool brown dwarfs \citep{Miles_2020}.  Figure \ref{fig:abundances_photochem} shows that CH$_4$ and CO quench at $\sim$ 10 bars in this best-fit case. This best-fit {\kzz} value is slightly inconsistent with the {\kzz} vs. $T_{\rm eff}$ trend observed in \citet{Miles_2020}, where {\kzz} values continue to be low at $T_{\rm eff}$ $>$ 400 K but shows a dramatic rise when $T_{\rm eff}$ $<$ 400 K. \citet{Mukherjee22} used atmospheric forward models with a self-consistent treatment of disequilibrium chemistry to theoretically explain this trend as a result of gases quenching in deep ``sandwiched" radiative zones with low {\kzz} in objects with 500 K $<$ $T_{\rm eff}$ $<$ 1000 K. The models showed that objects colder than 500 K tended to have gases quenched in their deep convective zones and are expected to show higher {\kzz} values representative of convective mixing. This theoretical trend was also found to have a significant gravity dependence in \citet{Mukherjee22} where objects with $\log(g$) $<$ 4.5 were expected to show convective zone quenching of gases across 400 K $<$ $T_{\rm eff}$ $<$ 1000 K. 

Given that our 5-knot retrievals show that our target has a $T_{\rm eff}$ of 458$^{+15}_{-15}$ K and $\log(g)$ of 4.46$^{+0.06}_{-0.04}$, our finding of a high {\kzz} makes it consistent with the trend predicted in \citet{Mukherjee22}. Therefore, it is likely that we are probing the deep convective zone {\kzz} in this object and not the radiative zone or ``sandwiched" radiative zone {\kzz}, as expected from self-consistent forward model trends. The maximum {\kzz} in the deep convective atmosphere of a brown dwarf with $T_{\rm eff}$ of 458 K and $\log(g)$=4.46 is 4.55$\times$10$^{10}$ cm$^2$s$^{-1}$, calculated using Equation 4 in \citet{Zahnle14}. This maximum {\kzz} in the convective zone is achieved when the entire energy flux from the interior is only carried out through convection in the deep atmosphere. However, in reality the interior energy flux is expected to be only partly carried out through convective transport and partly by radiative energy transport. In that case, the {\kzz} in the convective atmosphere is expected to be lower than this upper limit. Figure \ref{fig:abundances_photochem} also shows that our model fitting approach fits the abundances of all these gases quite satisfactorily except for NH$_3$. This might be suggestive of a slightly lower N/H ratio in the object than the scaled solar N/H ratio.

\subsection{Sensitivity to Thermal Profile Model}
\label{sec:MadhusudhanProfile}

In order to quantify whether our choice of thermal profile model impacts the resulting mixing ratios, we ran a second retrieval using the parametric thermal profile model described in \citet[][hereafter M\&S]{Madhusudhan_2009}.  In this model, the atmosphere is divided into three layers, for which the temperature and pressure are related by, 

\begin{align}
P_{0} < P < P_{1} &= P_{0}e^{\alpha_{1}(T-T_{0})^{1/2}} \, (\text{Layer} \: 1) \\
P_{1} < P < P_{2} &= P_{2}e^{\alpha_{2}(T-T_{2})^{1/2}} \, (\text{Layer} \: 2) \\
P_{3} < P &= T_{3} \, (\text{Layer} \: 3),
\end{align}

\noindent where $P_{0}$ and $T_{0}$ are the pressure and temperature at the top of the atmosphere, respectively. We eliminate the possibility of a thermal inversion in the atmosphere by setting $P_{2}$=$P_{1}$ and so we are left with 5 parameters: $\alpha_{1}$, $\alpha_{2}$, $P_{1}$, $P_{3}$, and $T_{3}$, the priors of which are given Table \ref{table:priors}.   In the appendix, Figure \ref{fig:Corner_prof2_complete} shows the marginalized posterior probability distributions for all 19 parameters and Table \ref{table:4} gives the median, and $1\sigma$ uncertainty for each of the parameters.

\begin{figure}[htb!]
    \centering
    \includegraphics[width=\linewidth]{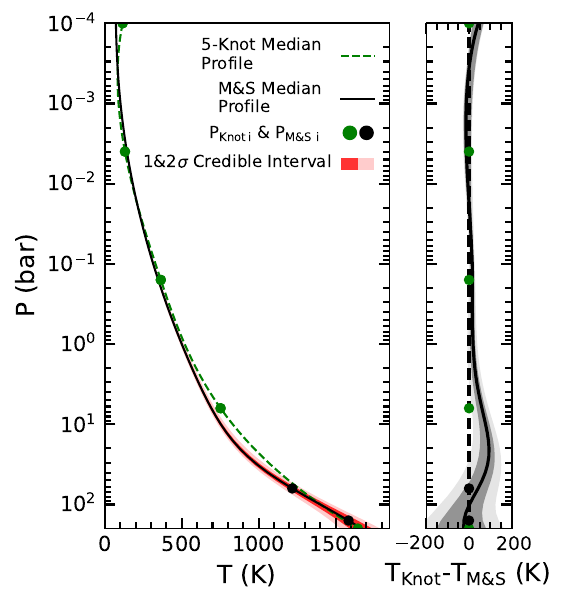}
    \caption{\textbf{Left panel}: shows the comparison between the two retrieved thermal profiles, where the black solid curve is the median M\&S thermal profile, the red region shows the 1 and 2$\sigma$ confidence interval around the median profile, and the green curve is the 5-knot median thermal profile. \textbf{Right panel}: shows the difference between the the 5-knot and the M\&S median profile, and the 1 and 2$\sigma$ central credible intervals, respectively.}
   \label{fig:T_P_compare}
\end{figure}

The left panel of Figure \ref{fig:T_P_compare} shows a comparison between the retrieved 5-knot thermal profile discussed in \S\ref{sec:5knotprofile} and the M\&S thermal profile while the right panel of Figure \ref{fig:T_P_compare} shows the differences between the two. The profiles agree within the uncertainties except below a pressure of a bar where the 5-knot profile is hotter by up to 100 K.

\begin{figure*}[htb!]
    \centering
    \includegraphics[width=\linewidth]{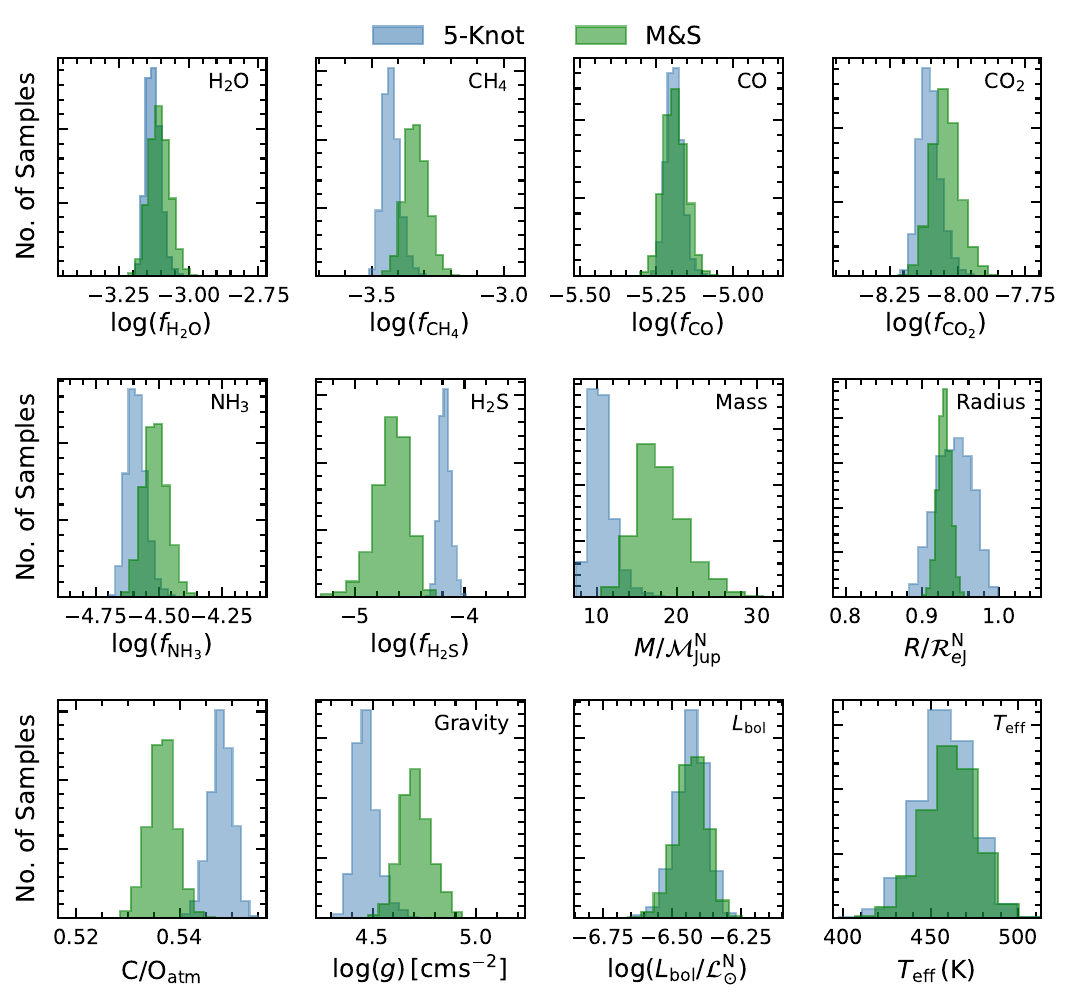}
    \caption{The posterior distributions retrieved for H$_2$O, CH$_4$, CO, CO$_2$, NH$_3$, H$_2$S, mass, and radius alongside the calculated properties like (C/O)$_\textrm{atm}$, gravity, $L_\textrm{bol}$, and $T_\textrm{eff}$ from the 5-knot (blue) and  M\&S (green) retrieval.}
   \label{fig:posteriors_compare}
\end{figure*}

Figure \ref{fig:posteriors_compare} shows a comparison between the posterior distributions of the 12 parameters shown in Figure \ref{fig:Corner_free} ($f_\mathrm{H_{2}O}$, $f_\mathrm{CH_{4}}$, $f_\mathrm{CO}$, $ f_\mathrm{CO_{2}}$, $f_\mathrm{NH_{3}}$, $f_\mathrm{H_{2}S}$, $M$, $R$, (C/O)$_\textrm{atm}$, $\log g$, $L_\textrm{bol}$, and $T_\textrm{eff}$)  from the 5-knot (blue) and the M\&S (green) retrieval. Overall the agreement between the distributions is good (see also Table \ref{table:4}) which suggests our results are not strongly dependent on the underlying thermal profile model.  The largest differences are for the distributions of $M$ and $\log (f_{\textrm{H}_2\textrm{S}})$ with fractional differences of the median values of $-0.64$ and $-0.10$, respectively.  

\begin{figure*}[htb!]
    \centering
    \includegraphics[width= \textwidth]{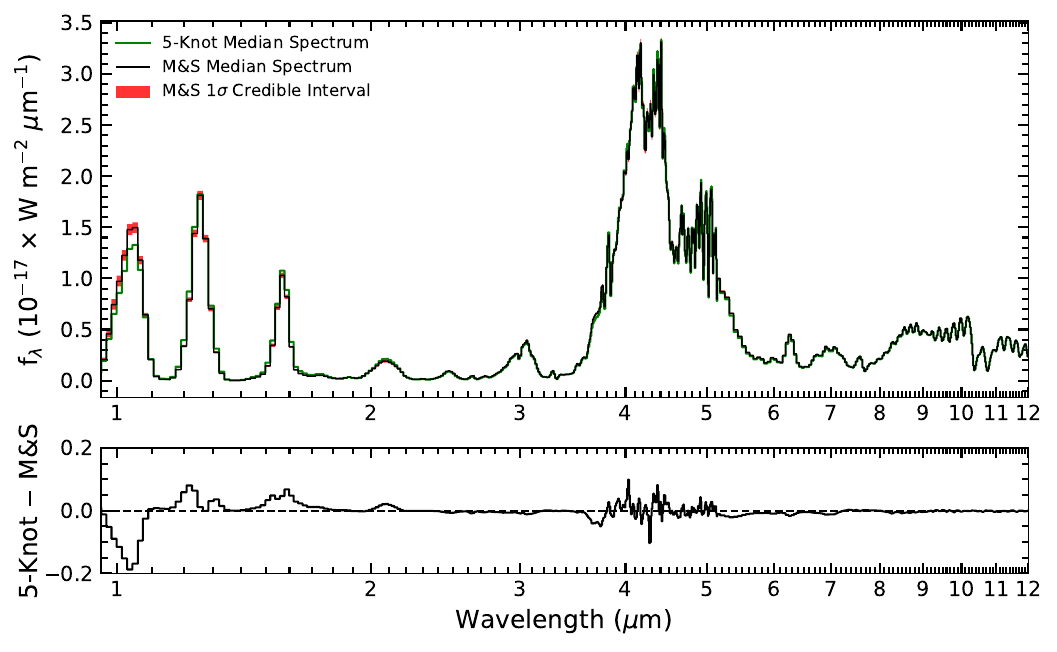}
    \caption{The top panel of the figure shows the retrieved median spectrum from M\&S retrieval in black and the retrieved median spectrum from 5-knot retrieval in green for WISE 0359--54 spectrum covering 0.96--12$\mu$m. The red region shows the 1$\sigma$ region around the M\&S median spectrum. The bottom panel of the figure shows the residual spectrum calculated by taking the difference between the retrieved M\&S median spectrum and the retrieved 5-knot median spectrum.}
   \label{fig:spec_parameterized}
\end{figure*}

Figure \ref{fig:spec_parameterized} shows the retrieved median model spectrum from the  M\&S retrieval (black) along with the 1$\sigma$ central credible interval (red) and the retrieved median model spectrum from the 5-knot retrieval (green); the lower panel shows the residual between the two models.  There 5-knot profile predicts systematically higher fluxes in the $J$-, $H$-, and $K$-band opacity holes at 1.25, 1.6, and 2.1 $\mu$m and systematically lower fluxes between 5 and 7 $\mu$m.  However, the median M\&S retrieval predicts a higher flux in the $Y$-band ($\sim$1 $\mu$m); this may be a result of the increased retrieved abundance of Na and K (0.26 and 0.04 dex larger, respectively) which forces light that would otherwise escape at wavelengths shorter than 1 $\mu$m to instead emerge in the $Y$-band opacity hole.

We chose nested sampling to sample posterior values due to its inherent ability to estimate the evidence, $p(\boldsymbol{D})$.  We can compute the posterior odds ratio between the 5-knot model and the M\&S model as,

\begin{equation}
\frac{p(\textrm{5-knot}|\boldsymbol D)}{p(\textrm{M\&S}|\boldsymbol D)} =   \frac{p(\textrm{5-knot})}{p(\textrm{M\&S})}  \hspace{0.05in} \frac{p(\boldsymbol D|\textrm{5-knot})}{p(\boldsymbol D|p(\boldsymbol{\Theta_\textrm{M\&S}))}}  
\end{equation}

\noindent
where the first term on the right-hand size is known as the prior odds and the last term on the right-hand side is known as the Bayes factor.  Assuming the prior odds ratio is unity, the posterior odds is simply given by the Bayes factor

\begin{equation}
B_\textrm{m} = \frac{p(\boldsymbol D|p(\boldsymbol{\Theta_\textrm{5-Knot}}))}{p(\boldsymbol D|p(\boldsymbol{\Theta_\textrm{M\&S}))}}.
\end{equation}

With ln $p(\boldsymbol D)$ values of $23540.66 \pm 0.37$ and $23560.97 \pm 0.36$ for the 5-knot and M\&S retrievals, respectively, we calculated a Bayes factor of 6.65 $\times$ 10$^{8}$.  Based on the Jeffreys' scale \citep{Jeffreys_1998}, this value suggest that the M\&S thermal profile is strongly preferred over the 5-knot profile.  We can convert this value to an equivalent ``$\sigma$" significance as described in  \citet{Benneke_2013} and find a value of 6.69$\sigma$.

\section{Summary}

In this work, we present an atmospheric retrieval analysis of the Y0 brown dwarf WISE 0359--54 using the low-resolution 0.96--12 $\mu$m JWST spectrum obtained using NIRSpec and MIRI. The cloudless retrieval was performed using the Brewster retrieval framework.  We retrieved volume number mixing ratios for 9 gases:  H$_{2}$O, CH$_{4}$, CO, CO$_{2}$, NH$_{3}$, H$_{2}$S, K, Na, PH$_{3}$. These retrieved mixing ratios are 3--5$\times$ more precise than the previous work done using the HST WFC3 data \citep{Zalesky_2019}. Since we were able to constrain all the major carbon- and oxygen-bearing molecules, we found (C/O)$_\textrm{atm}$ to be 0.548$\pm$0.002. Apart from constraining the chemical composition, we also found an order of magnitude improvement in the precision of the retrieved thermal profile, which can be attributed to the broad wavelength coverage of the JWST data.

Using the retrieved thermal profile and the calculated surface gravity, we generated a grid of forward models with varying metallicity [M/H], (C/O)$_\textrm{atm}$, and eddy diffusion coefficient (\kzz) which tells the atmospheric mixing vigor. Comparing these generated models with our retrieved mixing ratios of H$_{2}$O, CH$_{4}$, CO, CO$_{2}$ and NH$_{3}$, we found strong evidence of vertical mixing in the atmosphere of WISE 0359--54 with a value of $K_{zz}$=$10^9$ [cm$^{2}$s$^{-1}$]. 

Finally, to test the sensitivity of our results to our 5-knot thermal profile model, we performed another retrieval using the \citet{Madhusudhan_2009} thermal profile model. We found that the mixing ratios from both thermal profile model yield similar results (with the exception of $f_{\textrm{H}_2\textrm{S}}$ which is $-$0.10 dex lower) and that the retrieved thermal profile is similar except near the 5 bar pressure level where it is $\sim$100 K hotter.  Taken together, these results underscore the power that the James Webb Space Telescope has to study the atmospheres of the coolest brown dwarfs.

\section{Acknowledgement}

This work is based [in part] on observations made with the NASA/ESA/CSA James Webb Space Telescope. These observations are associated with program \#2302. 

This research has made use of the SIMBAD database, operated at CDS, Strasbourg, France.

We would also like to thank Jacqueline K. Faherty and 
Channon Visscher for their valuable insight on the retrieved results and the working of the chemistry in such cold objects.

\software{Corner \citep{corner_github},
Matplotlib \citep{matplotlib},
Numpy \citep{NumPy-Array}}

\newpage

\section{Appendix}

\begin{deluxetable*}{lccc}[htb!]
   \tablecaption{Posterior Parametric Values \label{table:4}}
   \tablehead{
   \colhead{Parameter} &
   \colhead{5-Knot Retrieval \tablenotemark{a}} & 
   \colhead{M\&S Retrieval \tablenotemark{a}} &
   \colhead{Fractional Difference \tablenotemark{b}}
   }
   \startdata
    $\log (f_\textrm{H$_{2}$O})$ & --3.13$^{+0.03}_{-0.02}$ & --3.10$^{+0.04}_{-0.04}$ & 0.01$^{+0.01}_{-0.01}$ \\ 
    $\log (f_\textrm{CH$_{4}$})$ & --3.43$^{+0.03}_{-0.03}$ & --3.34$^{+0.04}_{-0.04}$ & 0.03$^{+0.01}_{-0.01}$ \\
    $\log (f_\textrm{CO})$ & --5.19$^{+0.03}_{-0.02}$ & --5.18$^{+0.04}_{-0.04}$ & 0.00$^{+0.01}_{-0.01}$ \\
    $\log (f_\textrm{CO$_{2}$})$ & --8.11$^{+0.04}_{-0.03}$ & --8.05$^{+0.05}_{-0.05}$ & 0.01$^{+0.01}_{-0.01}$ \\
    $\log (f_\textrm{NH$_{3}$})$ & --4.59$^{+0.04}_{-0.03}$ & --4.51$^{+0.05}_{-0.05}$ & 0.02$^{+0.01}_{-0.01}$ \\
    $\log (f_\textrm{H$_{2}$S})$ & --4.18$^{+0.05}_{-0.05}$ & --4.60$^{+0.12}_{-0.12}$ & --0.10$^{+0.03}_{-0.03}$\phm{-}\phm{*} \\
    $\log (f_\textrm{K})$ & --9.31$^{+1.66}_{-1.68}$ & --8.96$^{+1.86}_{-1.96}$ & 0.03$^{+0.23}_{-0.28}$  \\    
    $\log (f_\textrm{Na})$ & --10.32$^{+1.08}_{-1.07}$\phm{*} & --7.45$^{+0.16}_{-0.49}$ & 0.26$^{+0.08}_{-0.11}$ \\
    $\log (f_\textrm{PH$_{3}$)}$ & --10.00$^{+1.15}_{-1.26}$\phm{*} & --10.22$^{+1.17}_{-1.13}$\phm{*} & --0.02$^{+0.15}_{-0.17}$\phm{-}\phm{*} \\ 
    $M/\mathcal{M}^{\mathrm{N}}_\mathrm{Jup}$ & \phm{-}10.40$^{+1.50}_{-1.10}$\phm{*} & 17.20$^{+2.80}_{-2.40}$ & --0.64$^{+0.30}_{-0.34}$\phm{-}\phm{*}  \\
    $R/\mathcal{R}^{\mathrm{N}}_{e\mathrm{J}}$ & \phm{-}0.94$^{+0.02}_{-0.02}$ & \phm{*}0.93$^{+0.01}_{-0.01}$ & \phm{-}0.01$^{+0.03}_{-0.03}$\phm{-} \\
    $\Delta \lambda$ & \phm{.}0.00$^{+0.00}_{-0.00}$ & \phm{.}0.00$^{+0.00}_{-0.00}$ & 0.04$^{+0.13}_{-0.14}$ \\
    log $b$ & --37.04$^{+0.04}_{-0.04}$\phm{*} & --37.02$^{+0.04}_{-0.04}$\phm{*} & 0.00$^{+0.00}_{-0.00}$  \\
    $d$ & --13.66$^{+0.33}_{-0.35}$\phm{*} & --13.50$^{+0.02}_{-0.03}$\phm{*} & 0.01$^{+0.01}_{-0.01}$  \\
    T$_\mathrm{Knot \, 1}$ & 115.71$^{+11.58}_{-13.26}$\phm{*}\phm{*} & $\cdots$ &  $\cdots$ \\
    T$_\mathrm{Knot \, 2}$ & 130.84$^{+11.06}_{-7.23}$\phm{*}\phm{*} & $\cdots$ &  $\cdots$\\
    T$_\mathrm{Knot \, 3}$ & 363.94$^{+2.80}_{-3.14}$\phm{*} & $\cdots$ &  $\cdots$\\
    T$_\mathrm{Knot \, 4}$ & 752.58$^{+7.72}_{-9.52}$\phm{*} & $\cdots$ &  $\cdots$\\
    T$_\mathrm{Knot \, 5}$ & 1734.47$^{+28.94}_{-34.37}$\phm{*}\phm{*} & $\cdots$ &  $\cdots$\\
    $\alpha_{1}$ & $\cdots$ & 0.45$^{+0.00}_{-0.00}$ &  $\cdots$\\
    $\alpha_{2}$ & $\cdots$ & 0.03$^{+0.01}_{-0.01}$ & $\cdots$ \\
    P$_{1}$ &$\cdots$ & 1.79$^{+0.10}_{-0.08}$ & $\cdots$ \\
    P$_{3}$ &$\cdots$ & 2.23$^{+0.06}_{-0.08}$ & $\cdots$ \\
    T$_{1}$ & $\cdots$ & 2048.75$^{+143.26}_{-161.03}$\phm{---}\phm{*}\phm{*} & $\cdots$ 
\enddata 

\tablecomments{\textsuperscript{a}All mixing ratios are reported as the log of the volume mixing ratio (the amount of molecular gas out of the total amount of molecular gas), where the remainder of the gas is assumed to be H\textsubscript{2}-He at a fixed solar ratio.\\
\textsuperscript{b} The difference between the 5-Knot and M\&S retrieved posterior samples divided by the 5-Knot retrieved posterior samples.}
\end{deluxetable*}

\begin{figure*}[htb!]
    \centering
    \includegraphics[width= \linewidth]{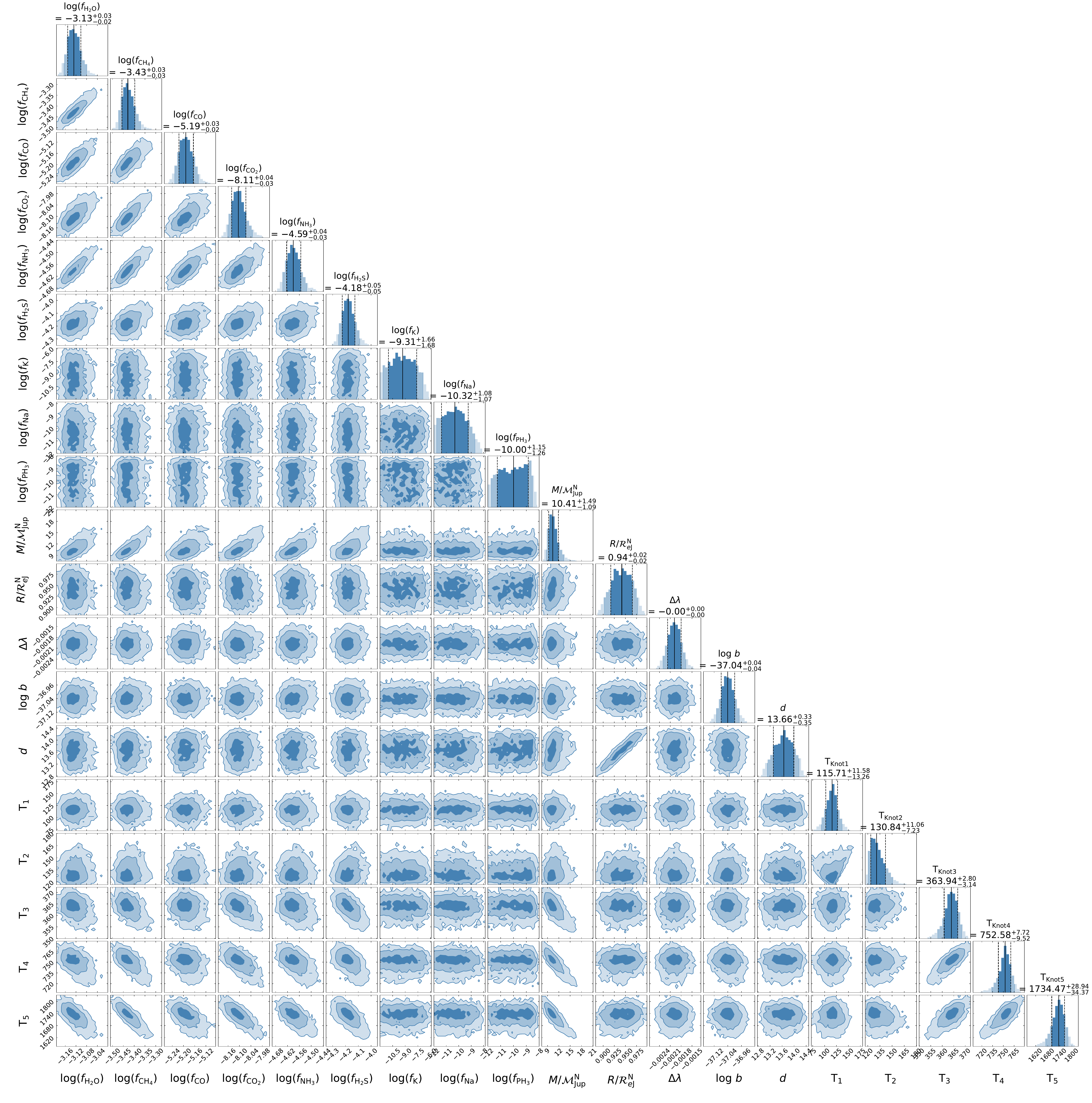}
    \caption{Marginalized posterior probability distributions for each parameter from the 5-knot retrieval for WISE 0359--54. The first 9 parameters represents the retrieved mixing ratios for H$_{2}$O, CH$_{4}$, CO, CO$_{2}$, NH$_{3}$, H$_{2}$S, K, Na and PH$_{3}$, followed by mass and radius. The parameters $\Delta \lambda$ and log $b$ are nuisance parameters, $d$ is the distance to the object, and the last five parameters are the retrieved temperature knots. The values above the 1--D histograms represents the parametric median (50th percentile) values with the errors representing the 1$\sigma$ central credible interval (16th and 84th percentile) values. The different shades in the 1--D and 2--D histograms represent the 1, 2 and 3$\sigma$ central credible interval, respectively, with the darkest shade corresponding to 1$\sigma$.}
   \label{fig:Corner_5_knot_complete}
\end{figure*}

\begin{figure*}[htb!]
    \centering
    \includegraphics[width= \textwidth]{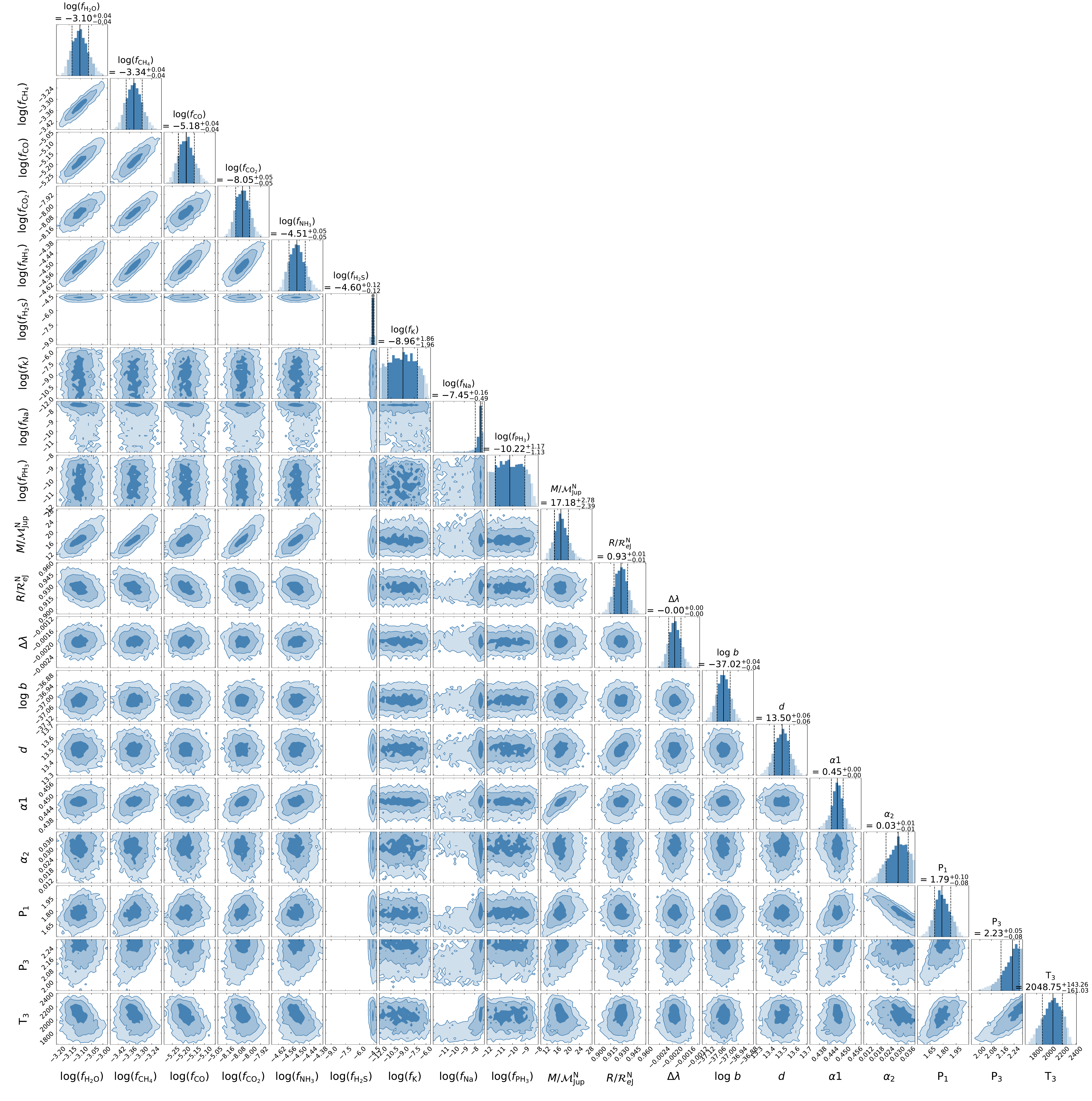}
    \caption{Marginalized posterior probability distributions for each parameter from the M\&S retrieval for WISE 0359--54. The first 9 parameters represents the retrieved mixing ratios for H$_{2}$O, CH$_{4}$, CO, CO$_{2}$, NH$_{3}$, H$_{2}$S, K, Na and PH$_{3}$, followed by mass and radius. The parameters $\Delta \lambda$ and log $b$ are nuisance parameters, $d$ is the distance to the object, and the last five parameters are the retrieved temperatures knots (points). The values above the 1--D histograms represents the parametric median (50th percentile) values with the errors representing the 1$\sigma$ central credible interval (16th and 84th percentile) values. The different shades in the 1--D and 2--D histograms represent the 1, 2 and 3$\sigma$ central credible interval, respectively, with the darkest shade corresponding to 1$\sigma$.}
   \label{fig:Corner_prof2_complete}
\end{figure*}

\newpage

\bibliographystyle{aasjournal}
\bibliography{Reference}

\end{document}